\documentstyle[a4,11pt]{article}
\begin{document}

\begin{titlepage}
\vskip 2cm
\begin{flushright}
Preprint CNLP-1994-05
\end{flushright}
\vskip 2cm
\begin{center}
 {\bf GAUGE EQUIVALENCE BETWEEN TWO-DIMENSIONAL HEISENBERG
FERROMAGNETS WITH SINGLE-SITE ANISOTROPY AND ZAKHAROV
EQUATIONS}\footnote{Preprint
CNLP-1994-05. Alma-Ata.1994 }
\end{center}
\vskip 2cm
\begin{center}
{\bf R. Myrzakulov }
\end{center}

\vskip 1cm
Centre for Nonlinear Problems, PO Box 30, 480035, Alma-Ata-35, Kazakhstan

E-mail: cnlpmyra@satsun.sci.kz

\vskip 1cm

\begin{abstract}
Gauge equivalence between the two-dimensional continuous classical
Heisenberg
ferromagnets(CCHF) of spin $\frac{1}{2}$-the M-I equation with single-side
anisotropy and the Zakharov equation(ZE) is
established for the easy axis case. The anisotropic CCHF is shown to be gauge
equivalent to the isotropic CCHF.
\end{abstract}

%\maketitle

\end{titlepage}

\setcounter{page}{1}
\newpage
\Large

In the study of 1+1 dimensional ferromagnets, a well
known Lakshmanan and gauge equivalence take place between the continuous classical
Heisenberg ferromagnets  of spin
$\frac{1}{2}$ and the nonlinear
Schrodinger equations (e.g., Lakshmanan 1977, Zakharov and Takhtajan 1979, Nakamura and
Sasaga 1982, Kundu and Pashaev 1983, Kotlyarov 1984). At the same time, there exit some integrable
analogues of the CCHF in 2+1 dimensions(Ishimori 1982, Myrzakulov 1987). One
integrable (2+1)-dimensional extension of the CCHF is the
following Myrzakulov-I(M-I) equation with one-ion anisotropy
$$
{\bf S}_t = ({\bf S} \wedge {\bf S}_{y} + u {\bf S})_{x} +
v {\bf S} \wedge {\bf n}, \eqno(1a)
$$
$$
u_{x} = - {\bf S} \cdot ({\bf S}_{x} \wedge {\bf S}_{y}),  \eqno(1b)
$$
$$ v_x =
\triangle ({\bf S}_y \cdot {\bf n}) \eqno(1c) $$
were the spin field $ {\bf S} = (S_1,S_2,S_3) $ with the magnetude normalized
to unity, $ u $ and $ v $ are scalar functions, $ {\bf n}=(0,0,1)$, and
$\triangle<0$ and $\triangle>0 $ correspond respectively to the system with
an easy plane and to that with an easy axis. Note that if the symmetry
$ \partial_x=\partial_y $ is imposed then the M-I equation (1) reduces
to the well known Landau-Lifshitz equation with single-site anisotropy
$$
{\bf S}_t = {\bf S} \wedge ({\bf S}_{xx} +
\triangle ({\bf S} \cdot {\bf n}){\bf n}). \eqno(2)
$$

The aim of this letter is the construction  the (2+1)-dimensional nonlinear
Schrodinger equation which is gauge equivalent
to the M-I equation (or two-dimensional CCHF)(1) with the easy-axis
anisotropy $ (\triangle > 0). $ Besides the gauge equivalence between
anisotropic
$ (\triangle \not= 0) $ and isotropic $ (\triangle = 0 ) $ CCNF (1) is
established. The Lax representation of the M-I equation (1) may be given by
(Myrzakulov 1987)
$$ \psi_x=L_1 \psi,\,\,\,\, \psi_t=2\lambda \psi_y+M_1 \psi \eqno (3) $$
where
$$ L_1=i \lambda S+ \mu[\sigma_3, S],\,\,\,\,\,
M_1=2\lambda A+2i\mu[A,\sigma_3]+4i \mu^2 \{\sigma_3, V\}\sigma_3 \eqno(4)$$
with $$
S=\sum_{k=1}^3 S_k \sigma_k,\,\,\,\, V=\triangle\partial^{-1}_{x} S_y dx,\,\,\,\,\,
A=\frac{1}{4}([S,S_y]+2iuS), \,\, \mu=\sqrt\frac{\triangle}{4}, \triangle >0.
$$
Here $ \sigma_k $ is Pauli matrix, [,] (\{,\}) denote commutator
(anticommutator)
and $\lambda $ is a spectral parameter. The matrix $ S $ has the following
properties: $ S^2=I, \,\,\, S^{\ast} = S,\,\,\, trS=0 $.
The compatibility condition of system (2) $ \psi_{xt}=\psi_{tx} $ gives the
M-I equation (1).
Let us now consider the gauge transformation induced by $ g(x,y,t): \psi=
g^{-1}\phi$, where $g^{\ast}=g^{-1}\in SU(2)$. It follows from the properties
of the matrix S that it can be representies in the form $ S=
g^{-1}\sigma_3g$. The new
gauge equivalent operators $ L_2, M_2 $ there fore should be given by
$$ L_2=gL_1g^{-1}+g_xg^{-1},\,\,\,\, M_2=gM_1g^{-1}+g_tg^{-1}
- 2\lambda g_{y}g^{-1}  \eqno(5) $$
and satisfy the following system of equations
$$ \phi_x=L_2\phi,\,\,\,\,\, \phi_t=2\lambda\phi_y+M_2\phi \eqno(6) $$
Now choosing
$$ g_xg^{-1} + \mu g[\sigma_{3},S]g^{-1}=U_{0},\,\,\,gSg^{-1}=\sigma_{3},
 \eqno(7a) $$
$$ g_t g^{-1}+2i\mu g[A,\sigma_3]g^{-1}+
4i\mu^2 g \{\sigma_3,V\}\sigma_3g^{-1}=V_{0}, \eqno(7b)
$$
with
$$
U_{0} =
\left ( \begin{array}{cc}
0       & q \\
-\bar q & 0
\end{array} \right),\,\,\,\,\,
V_{0} =
i\sigma_{3}(\partial^{-1}_{x}|q|^{2}_{y} - U_{0y}).
$$
where $ q(x,y,t) $ the new complex valued fields. Hence we finally obtain
$$ L_2=i\lambda \sigma_3+U_{0},\,\,\,\,\, M_2=V_{0} \eqno(8) $$
The compatibility condition $ \phi_{xt}=\phi_{tx} $ of the system (6) with
the operators $ L_2, M_2$ (8) leads to the (2+1)-dimensional Zakharov
equation (Zakharov 1979, Strachan 1993)
$$
iq_t=q_{xy}+wq,\,\,\,\, w_x =2(|q|^2)_y \eqno(9)
$$

We note that under the reduction $ \partial_{y} = \partial_{x} $
equation(9) becomes the well known
(1+1)-dimensional NLSE. Thus we have shown that the M-I equation with
single-site anisotropy (the two dimensional continuum Heisenberg
ferromagnets) is gauge equivalent to the (2+1)-dimensional NLSE - the
Zakharov equation(9). On the other hand, equations (9) is gauge and
geometrical equivalent to the isotropic M-I equation ( Myrzakulov 1987;
Nugmanova 1992, Myrzakulov  1994)
$$ iS^{\prime}_t=\frac{1}{2}([S^{\prime}, S^{\prime}_y]+2iu^{\prime}
S^{\prime})_x \eqno(10a) $$
$$ u^{\prime}_x+{\bf S^{\prime}}({\bf S^{\prime}}_x\wedge
{\bf S^{\prime}}_y)=0 \eqno(10b) $$
which is the compatibility condition of the following system
$$ f_x = L^{\prime}_{1} f,\,\,\,\, f_{t} =2\lambda f_y+
\lambda M^{\prime}_{1}f \eqno(11)
$$
where
$$ L^{\prime}_1=i\lambda S^{\prime},\,\,\,\,\, M^{\prime}_1=\frac{1}{2}
([S^{\prime},S^{\prime}_y]+2iuS^{\prime}). \eqno(12) $$

Now we show below the gauge equivalence between anisotropic and isotropic M-I
equations (1) and (10),respectively. Indeed the Lax representations (3) and (11),
which reproduce equations (1) and (10) respectively can be obtained from
each other by the $ \lambda $-independent gauge transformation $ h(x,y,t)=
\psi^{-1}|_{\lambda=b} $ as
$$ L_{1}^{\prime}=hL_{1}h^{-1}+h_{x}h^{-1},\,\,\,\,\,\,
M^{\prime}_{1}=hM_{1} h^{-1}+h_{t} h^{-1}. \eqno(13)
$$

In this way, the solutions of eq.(1) and (10) connected each other by formulas
$ S=h^{-1} S^{\prime}h. $ Now we present the important relations between
field
variables $ \psi $ and S
$$
|\psi|^2=\frac{1}{2}[{\bf S}^{2}_{x}-8\mu S_{3x}+16\mu^{2}(1-S^{3}_{2})]
\eqno(14a)
$$
$$
\psi\bar\psi_{x} - \bar\psi \psi_{x}=
\frac{i}{4}{\bf S}\cdot{[{\bf S_{xx}}+16\mu^2({\bf S}\cdot
{\bf n}){\bf n})]+4\mu{\bf S}\cdot({\bf S}_{xx}\wedge {\bf n})} \eqno(14b) $$

These relations coincide with the corresponding connections between
$ q $ and $S$ of one-dimensional case(Guispel and Capel 1982, Nakamura
and Sasada 1982).

Note that the two-dimensional CHSC with ($\triangle < 0$)
easy
plane single-site anisotropy is gauge equivalent to the some general (2+1)-
dimensional NLSE (Myrzakulov 1987):
$$
iq_t=q_{xy}+wq \eqno(15a)
$$
$$
ip_t=-p_{xy}-wp \eqno(15b)
$$
$$
w_x =2(pq)_y \eqno(15c)
$$
Besides, the two-dimensional CHSC, when
$ S \in SU(1,1)/U(1)$, i.e. the non-compact group, is gauge and Lakshmanan
equivalent to the Zakharov equation(15) with the repulsive interaction,
$p=-\bar q$ (Myrzakulov 1987).

Finally we note that the M-I equation(1) is the particular case of the
M-III equation(Myrzakulov 1987)
$$ {\bf S}_{t}=({\bf S}\wedge {\bf S}_{y}+u{\bf S})_x+2b(cb+d){\bf S}_{y}
     -4cv{\bf S}_{x} + {\bf S} \wedge {\bf V} \eqno (16a) $$
$$ u_x=-{\bf S}\dot({\bf S}_{x}\wedge {\bf S}_{y}), \eqno(16b)$$
$$ v_x=\frac{1}{4(2bc+d)^2}
   ({\bf S}^{2}_{1x})_y \eqno (16c) $$
$$ {\bf V}_x= J{\bf S}_{y}, \eqno(16d)$$
where $J$ is a anisotropic matrix. These equations admit the some integrable reductions.

To summarise, we have established the gauge equivalence
between the classical continuum Heisenberg spin $ \frac{1}{2} $ chain with
easy axis ($\triangle >0$) anisotropy and the nonlinear Schrodinger equation
of the attractive type in two-dimensions. In additional the anisotropic
CHSC(1) is shown to be gauge equivalent to isotropic CHSC(10).
These results corresponds to the physical statements on thermodynamical equivalence between the
corresponding quantum versions of Heisenberg ferromagnet(antiferromagnet) and Bose gas
with attractive(repulsive) interaction, respectively(Yang and Yang 1966).

\end{document}